\documentstyle[preprint,eqsecnum,prb,aps]{revtex}
\begin{document}
\draft

\preprint{NSF-ITP-93-76}

\title{Kohn Anomalies in Superconductors}

\author{Michael E. Flatt\'e}
\address{Institute for Theoretical Physics,
University of California,
Santa Barbara, CA 93106-4030}
\date{May 14, 1993}
\maketitle
\begin{abstract}
I present the detailed behavior of phonon dispersion curves near momenta
which span the electronic Fermi sea in a superconductor.
I demonstrate that
an anomaly, similar to the metallic Kohn anomaly, exists in a
superconductor's dispersion curves when the frequency of the phonon
spanning the Fermi sea exceeds
twice the superconducting energy
gap. This anomaly occurs at approximately the same momentum
but is {\it stronger} than the normal-state Kohn anomaly. It also survives at
finite temperature, unlike the metallic anomaly.
Determination of
Fermi surface diameters from the location of these anomalies,
therefore,
may be more successful in the superconducting phase than in
the normal state. However, the superconductor's anomaly
fades rapidly with increased phonon frequency and becomes
unobservable when the phonon frequency greatly exceeds the gap. This
constraint makes these anomalies useful only in high-temperature
superconductors such as $\rm La_{1.85}Sr_{.15}CuO_4$.
\end{abstract}
\narrowtext

\section{Introduction}

The Kohn anomaly\cite{Kohn} occurs in a metal's phonon dispersion
curves when a phonon's momentum spans the Fermi surface. Locating
these anomalies through inelastic neutron scattering (on lead\cite{lead}
or niobium,\cite{niobium} for
example) and inelastic helium scattering (on
a platinum surface\cite{platinum}), accurately measures
the Fermi surface, as well as the electron-phonon interaction.
This Article consists of a derivation and discussion of a
similar type of anomaly, with greater magnitude, which exists in a
superconductor.  This anomaly could prove useful in
$\rm La_{1.85}Sr_{.15}CuO_4$, whose Fermi surface shape generates
heated debate.

A significant decay product of a phonon in a metal
is a single electron-hole pair.
The Kohn anomaly occurs
because, for momenta smaller than the Fermi surface diameter, there exist
single-pair excitations of
the electron gas for the phonon to decay into, while for
larger momenta there are none. This sharp change in the availability of decay
products causes a nonanalyticity in the phonon's lifetime and, by a
Kramers-Kronig relation, in its frequency. The sharpness originates in
the discontinuous electron occupation at the Fermi
surface at $0$K.  Thus, even in an interacting electron gas, with a
quasiparticle weight less than unity, the anomaly persists.
The discontinuity vanishes at finite temperature,
resulting in a phonon anomaly smoothed over the
momentum range $k_BT/\hbar v_F$, where $v_F$ is the Fermi velocity.
This smoothing is typically unobservable. However, in the high-temperature
superconductor $\rm La_{2-x}Sr_{x}CuO_4$
at room temperature any Kohn anomaly would be substantially
smoothed.
This may explain the failure of a search\cite{roomtempKA}
for Kohn anomalies in that material.
In order to lay the foundation for discussion of the Kohn anomalies in
superconducting
$\rm La_{1.85}Sr_{.15}CuO_4$,
this Article begins with the characteristics of that material's {\it metallic}
Kohn anomalies.

A standard
approximation in the derivation of the metallic anomaly is the
use of the {\it static}
pair response function.  Neglecting the phonon frequency
is suggested by the smoothness of the metal's
electronic response function at
frequencies much smaller than the Fermi energy $\epsilon_F$.  This
smoothness persists at finite temperature, also
justifying a
static approximation. In most superconductors, however,
the energy gap $\Delta$ produces
substantial structure in the response function at frequencies much less
than
phonon frequencies ($\Delta\ll \hbar\omega_D$).
Thus the static electronic response differs qualitatively from that at a
phonon's frequency.

In work primarily devoted to calculating the screening around a static
impurity, Hurault\cite{Hurault}
suggested that a superconductor has no true Kohn anomaly. Since
even at $0K$ the
electronic occupation is continuous in
the superconductor\cite{Schrieffer}, he argued that
the metallic anomaly would appear
smoothed over a momentum range equal to the inverse coherence length
$\hbar/\xi =\Delta/\hbar v_F$. Testing this prediction proved impossible
since the momentum resolution of inelastic neutron scattering
would not suffice. For high-temperature superconductors,
however, since $\hbar /\xi\sim .1\AA^{-1}$ the resolution is
adequate.

A heuristic explanation for the survival of Kohn anomalies in phonon
dispersion curves whose frequencies exceed $2\Delta$ follows.
Fig. 1a shows the minimum-energy electronic excitations
(from now on, the adjective ``single-pair'' will be dropped)
for a two or three-dimensional isotropic-gap
superconductor (solid line) and normal metal
(dashed line).
In Fig. 1b the region near $q=2k_F$ has been enlarged
so that the solid and
dashed lines can be distinguished.
For the
superconductor, in the region to the left of and above the solid line
there exist excitations, so the
electronic
response function has a finite imaginary component.
To the right of and below the solid line,
however, no excitations exist, so the  response function is real.  A
function must be  nonanalytic on the border between a region
where  it is identically zero and a region where it is nonzero;
the imaginary part of the response function is nonanalytic on
this (solid) line.
By Kramers-Kronig relations, the real part is nonanalytic there as well.
Thus the superconductor must produce an anomaly in phonon dispersion
curves at $q\sim 2k_F$
when $\hbar\omega(2k_F)>2\Delta$. All phonons resolvable by neutron
scattering in low-temperature
superconductors satisfy this condition as do most in
high-temperature ones.

For $2\Delta\alt\hbar\omega$, enhancement of the density of states
near the Fermi surface due to superconductivity enlarges the
superconductor's anomaly relative to that of the metal.
The different character of the large-momentum excitations in a
superconductor also augments the anomaly for $2\Delta\alt\hbar\omega$.
However, for phonon energies $\hbar\omega$
far above $2\Delta$, the altered character of large-momentum excitations
renders the anomaly
unobservable. The observability condition reduces to the resolvability
of $\xi^{-1}$. For this reason,
only high-temperature superconductors may have observable differences in
their Kohn anomalies between the metallic and superconducting phases.

Figs. 1ab illuminate the two types of anomalies which occur in a
superconductor.
Anomalies in phonon dispersion curves in superconductors
when $\hbar\omega = 2\Delta$
occur for zero momentum up to the Fermi surface
diameter.
They were proposed by Bobetic,\cite{Bobetic} elaborated by
Schuster,\cite{Schuster}  and observed in
$\rm Nb_3Sn$\cite{nb3sn}
and niobium\cite{niobsc}.
 Recently\cite{Flatte} it has been pointed out that in
quasi-two-dimensional superconductors with anisotropic gaps, the
frequency where these anomalies occur depends on the phonon momentum.
This observation forms the basis of a method
for measuring the energy-gap anisotropy in a high-temperature
superconductor.
The remainder of this Article will consider the
anomalies induced in phonon dispersion curves crossing the
solid line when $q\sim 2k_F$.

It is important to note that numerical calculations of the effect of
d-wave and s-wave superconductivity on phonon lifetimes and frequencies
have been performed for a nearest-neighbor tight-binding
model\cite{Marsiglio}. It is possible to find features in these results
which resemble the Kohn anomalies to be discussed in this article.
However, the results presented here concern the location and analytic
form of the anomalies, which were not discussed in Ref 13.
Furthermore, the primary concern of Ref. 13 was to locate
features identifying nesting, or which
distinguish s-wave from d-wave gaps. I do not address nesting because
the appearance of Kohn anomalies
does not depend on nesting, merely the diameter of the Fermi surface.
And, as I discuss in
Section V, the analytic form of the Kohn anomalies is identical for
s-wave and d-wave gaps for almost all phonon momenta.

\section{Normal State}

The conditions outlined in Section I for the observation of
a Kohn anomaly in a superconductor ($\xi^{-1}$ resolvable by neutron
scattering and $2\Delta\alt \hbar\omega$) imply that
$\hbar\omega/\epsilon_F$ is of order $10^{-1}$. This has implications
for the momentum of the Kohn anomaly in the metal.

The metal's electron-hole response function,
\begin{equation}
P_{M}({\bf  q},\omega) = \lim_{\alpha\to 0}
\sum_{{\bf
k}}{f(\epsilon_{\bf k}) - f(\epsilon_{{\bf k}+{\bf q}})\over
\epsilon_{{\bf k}} - \epsilon_{{\bf k}+{\bf q}} - \hbar\omega -
i\alpha},\label{nmresp}
\end{equation}
depends on the ratio $\hbar\omega/\epsilon_F$. Here $f(\epsilon)$ is the
Fermi function and $\epsilon_{\bf k}$ the dispersion relation for the
metal's electrons.
The response in Eq. (\ref{nmresp})  is called
the Lindhard function for a spherical Fermi surface in three
dimensions.
\begin{eqnarray}
&& P_M^{\rm 3D}({\bf  q},\omega) = -{N^*\over 4x}\bigg(2x -
{(1-[x+\nu]^2)\over 2}{\rm
ln}\left| {1-x-\nu\over 1+x+\nu}\right| \nonumber\\
&&\qquad\qquad - {(1-[x-\nu]^2)\over 2}{\rm
ln}\left| {1-x+\nu\over 1+x-\nu}\right| \nonumber\\ &&-
{i\pi\over 2}\left[\theta(1-|x+\nu|)(1-[x+\nu]^2)-\theta
(1-|x-\nu|)(1-[x-\nu]^2)\right]\bigg)\label{3dpi}
\end{eqnarray}
where $x=q/2k_F$, $\nu= m\omega/\hbar qk_F$, $N^*$ is the density of
states at the Fermi surface, and $\theta$ is the Heavyside step
function.
In the limit $\hbar\omega/\epsilon_F\to 0$,
\begin{equation}
 P_M^{\rm 3D}({\bf  q},0) = -{N^*\over 4x}\left(2x - {(1-x^2)}{\rm
ln}\left| {1-x\over 1+x}\right| \right).
\label{3dpiw=0}
\end{equation}
The location of the Kohn anomaly at the momentum $2k_F$ follows directly
from the nonanalyticity of the right hand side of Eq. (\ref{3dpiw=0})
at that momentum.
Clearly from Eq. (\ref{3dpi}), however, at finite frequency the nonanalyticity
takes place at
\begin{equation}
q_n=k_F\pm k_F\sqrt{1-{\hbar\omega\over\epsilon_F}} \sim 2k_F\pm {k_F\over
2}{\hbar\omega\over \epsilon_F}.\label{loca}
\end{equation}

Fig.~\ref{types} indicates the four types of extremal excitations which
produce anomalies in the response function. Excitation (1) takes an
electron from the Fermi surface and places it in a state $\hbar\omega$
above the Fermi surface on the other side of the Fermi sea. Excitation
(3) takes an electron from a state $\hbar\omega$ below the Fermi surface
and places it on the Fermi surface on the other side. For momenta
greater than (1) or less than (3) these types of excitations do not
exist.
This explains the origin of the two
solutions for $q_n$. For $\hbar\omega\approx 0$, excitations (1) and (3)
are the same.
The static
approximation succeeds because the nonanalyticity has the same form for
finite frequency as for zero frequency
and because the differences in $q_n$ cannot be resolved. Excitations (2)
and (4) concern the zero-momentum anomaly in a metal's response
function which will not be discussed in this Article.

Fig.~\ref{3dmetal}
shows $ P_M^{\rm 3D}({\bf  q},\omega)$ for various values of
$\hbar\omega/\epsilon_F$. These are plotted to indicate the changes in
the anomalies' momenta due to finite frequencies.
In a high-temperature superconductor, where the bandwidth may be less than
an electron volt and the phonon energies are tens of meV, the
splitting evident in Eq. (\ref{loca}) may be observable.

Another feature of the high-temperature superconductors is that their
electronic structure is quasi-two-dimensional.
In two dimensions the slope of the response function\cite{Stern}
is discontinuous and
divergent at $q_n$:
\begin{eqnarray}
&&P_M^{\rm 2D}({\bf  q},\omega)=-{N^*\over 2x}
\bigg(2x-{\rm sgn}(x+\nu)\theta(|x+\nu|-1)\sqrt{(x+\nu)^2-1}\nonumber\\
&&\qquad\qquad -
{\rm sgn}(x-\nu)
\theta(|x-\nu|-1)\sqrt{(x+\nu)^2-1}\nonumber\\ &&-
i\theta(1-|x+\nu|)\sqrt{1-(x+\nu)^2}+i\theta(
1-|x-\nu|)\sqrt{1-(x-\nu)^2}\bigg). \label{2dpi}
\end{eqnarray}
The two-dimensional response
contains stronger nonanalyticities than the three-dimensional response.
Fig.~\ref{2dmetal} shows $P_M^{\rm 2D}({\bf  q},\omega)$ for various values of
$\hbar\omega/\epsilon_F$.

Figs.~\ref{100KA} and \ref{110KA} indicate the location
of Kohn anomalies in $({\bf
q},\omega)$ space in the (100) and (110) directions for
$\rm La_{1.85}Sr_{.15}CuO_4$, using the
Fermi surface parametrizations of Hybertsen {\it et al.}\cite{Hybertsen}. The
low-energy phonons are also plotted as the solid
lines\cite{Pintschovius}.
Every time a dispersion curve crosses one of these lines,
a Kohn anomaly should appear. In the (100) direction the difference in
momentum between the actual anomaly and the static anomaly may be
visible in high-energy phonons.
Unfortunately, a recent experiment\cite{roomtempKA}
looking for Kohn anomalies in $\rm La_{1.9}Sr_{.1}CuO_4$
was performed at temperatures too high to
see this splitting ($k_BT>\hbar\omega$) and probably too high
($k_BT/\epsilon_F\sim 0.1$)
to see anomalies at all.

\section{Kohn Anomalies in Superconductors}

\subsection{$\hbar\omega<2\Delta$}

The disappearance of the Kohn anomaly in a superconductor was
suggested by Hurault\cite{Hurault}
 as a manifestation of
Fermi surface smoothing in a superconductor.  To explain this requires
formal machinery. The quasiparticle description, due to  Bogoliubov,
provides the most
convenient method of calculating the
effect of the superconducting
electron system on the phonons\cite{Schrieffer}.
The quasiparticle creation and
annihilation operators $\gamma$ relate to the electron creation
and annihilation operators $c$ as follows:
\begin{eqnarray}\gamma_{{\bf k},\uparrow}^{\dag} &&= u_{\bf k}c_{{\bf
k},\uparrow}^{\dag} - v_{\bf k}c_{-{\bf k},\downarrow}\nonumber\\
\gamma_{-{\bf k},\downarrow} &&= u_{\bf k} c_{-{\bf
k},\downarrow} + v_{\bf k}c_{{\bf k},\uparrow}^{\dag}.\label{qp}
\end{eqnarray}
Here
\begin{equation}
u_{\bf k} = {1\over \sqrt{2}}\left(1+{\epsilon_{\bf k}\over
E_{\bf k}}\right)^{1\over 2},\qquad
v_{\bf k} = {1\over \sqrt{2}}\left(1-{\epsilon_{\bf k}\over
E_{\bf k}}\right)^{1\over 2},\qquad u_{\bf k}v_{\bf
k} = {\Delta\over 2E_{\bf k}},\label{usvs}
\end{equation}
where
$E_{\bf k} =
\sqrt{\epsilon^2_{\bf k} + \Delta^2 }$ is the energy added to the
system by creating a quasiparticle of momentum ${\bf k}$.  The
Hamiltonian, expressed in quasiparticle operators, is then
\begin{equation}
H_o = \sum_{{\bf k},s}E_{{\bf k}}\gamma^{\dag}_{{\bf
k},s}\gamma_{{\bf k},s}\label{hamil}
\end{equation}
and the ground state contains no quasiparticles.

A significant difference between the superconducting system and a
normal system is the $v_{\bf k}$ function, which
is the analogue of the Fermi occupation function $f(\epsilon_{\bf
k})$ in a metal.  At zero temperature $f(\epsilon_{\bf k})$
has a discontinuity at the Fermi momentum, while
$v_{\bf k}$ smoothly
falls to zero over a momentum range $(\hbar/\xi)$.
$v_{\bf k}$ and $f(\epsilon_{\bf k})$ are shown in Fig.~\ref{vpfe}.
The Kohn anomaly arises from
the discontinuity in the electron occupation, and
as $f(\epsilon_{\bf
k})$ becomes smoother due to increased temperature, the apparent anomaly
becomes
weaker\cite{BealMonod}. Hurault
suggested the smoothness
of $v_{\bf k}$ due to superconductivity affected the Kohn
anomaly the same way as the smoothness of $f(\epsilon_{\bf
k})$ at finite
temperature affected the anomaly.  He
predicted a ``smoothing''
of the Kohn anomaly over a momentum range of $(\hbar/\xi)$ and extracted
this smoothing from the superconductor's static response function.

However, this heuristic explanation needs to be reexamined in the light
of the
existence of anomalies for higher phonon frequencies.
The Fermi surface sharpness cannot change as a function
of frequency in the superconductor.  Instead, the explanation for the
smoothing of the Kohn anomaly at small frequencies must be due to the
lack of {\it any} electronic excitations in the superconductor,
at small {\it or} large momenta (as seen in Figs. 1a and 1b).

Calculating the non-analytic behavior of an anomaly is necessary
to make this argument concrete. This calculation requires
the superconducting electronic response function,
\begin{equation}
P_S({\bf q},\omega) =  \lim_{\alpha\to 0}\sum_{{\bf k}} \left({
E_{\bf k}E_{{\bf q}-{\bf k}} -
\epsilon_{\bf k}\epsilon_{{\bf q}-{\bf k}} + Re(\Delta_{\bf k}
\Delta^*_{{\bf q} - {\bf k}})\over
2E_{\bf k}E_{{\bf q}-{\bf k}}}\right){-1\over E_{\bf k}+E_{{\bf q}-{\bf
k}}-
\hbar\omega-i\alpha},\label{Psc}
\end{equation}
where the sum is over {\it all} $\bf k$ values.
The parenthetical factor, called the coherence factor and denoted
$C({\bf k},{\bf q}-{\bf k})$,
reaches its maximum of
$1$ for $\bf k$ and ${\bf q}-{\bf k}$ at the Fermi surface.
It behaves similarly to the occupation expression
$f(\epsilon_{\bf k})-f(\epsilon_{{\bf k}+{\bf q}})$
in the normal metal's response function, Eq. (\ref{nmresp}),
but
while the occupation expression vanishes sharply as a function of
$\bf k$  or $\bf q$, $C$ decays to zero  with a scale given
by the inverse coherence length.  For the rest of this section, the gap
will be assumed isotropic. Section V will discuss anisotropic gaps.

The second factor of Eq. (\ref{Psc}),
the energy denominator, has poles for all excitations of the
Bogoliubov quasiparticle sea.
The imaginary
part of $P_S({\bf q},\omega)$ consists of
contributions from each of these poles (the
coherence factor is real).

At zero frequency there are no excitations in the
isotropic superconductor. Since there are no poles of the energy
denominator, $P_S({\bf q},\omega)$ is real for all $\bf q$.
The smoothness of the integrand in
Eq. (\ref{Psc}) with respect to $\bf q$ for all values of $\bf k$
forces the response function to be smooth with respect to
$\bf q$.

This smoothness can be estimated in a simple way from the change of
$P_S({\bf q},\omega)$ at $q=2k_F$.
In three
dimensions the sum from Eq. (\ref{Psc}) can be replaced by the
following integral:
\begin{equation}
P_S^{\rm 3D}({\bf  q},0)  = -\left({m\over 2qk_F\hbar^2}\right)N^*
\int_{-\epsilon_F}^{\infty}d\epsilon\int_{{\hbar^2(k-q)^2\over
2m}}^{{\hbar^2(k+q)^2\over 2m}}d\epsilon'{EE'-\epsilon\epsilon'
+\Delta^2\over 2EE'(E+E')}.\label{3dsco}
\end{equation}
Here $E=\sqrt{\epsilon^2+\Delta^2}$ and $N^*$
is the density of states
per unit energy at the Fermi energy in an otherwise
identical material with $\Delta=0$.  This usually is the
normal metal.
The integral in Eq. (\ref{3dsco}) can be estimated
near $q=2k_F$ to yield a measure of the remnant of the anomaly:
\begin{equation}
2k_F{\partial P_S^{\rm 3D}({\bf  q},0) \over \partial q}
\bigg|_{2k_F}\bigg/P_S^{3D}(2{\bf
k_F},0)\sim
2\ln\left({\Delta\over \epsilon_F}\right)\sim
2\ln\left({2\over k_F\xi}\right).\label{remnant3d}
\end{equation}
When $\Delta$ vanishes, the logarithmically-divergent slope
of the normal-metal response reemerges.  That response, the
Lindhard function, is Eq. (\ref{3dpi}).

For finite
but small frequencies in the superconductor,
the slope magnitude
increases to
\begin{equation}
\left|2k_F{\partial  P_S^{\rm 3D}({\bf  q},\omega) \over
\partial q}\bigg|_{2k_F} \bigg/P_S^{3D}(2{\bf
k_F},0)\right|\sim
\ln\left({\epsilon_F^2\over \Delta^2-(\hbar\omega/2)^2 }\right).\label{remff3d}
\end{equation}
This increase results from the overall decrease in all the
energy denominators in Eq. (\ref{Psc}),
due to a finite driving frequency. That
change
increases the contribution of each virtual excitation to the
response of the superconductor.
The overall response of
the superconductor also increases, so the relative magnitude
of the slope does not change for small but finite frequency.
\begin{equation}
\left|2k_F{\partial
 P_S^{\rm 3D}({\bf  q},\omega) \over \partial q}\bigg|_{2k_F}
\bigg/P_S^{3D}(2{\bf
k_F},\omega)\right|\sim
2\ln\left({\Delta\over \epsilon_F}\right)\sim
2\ln\left({2\over k_F\xi}\right).\label{remnanw3d}
\end{equation}

In a quasi-two-dimensional superconductor a similar effect occurs.
Instead of diverging as in Eq. (\ref{2dpi}),
however, the slope magnitude reaches a maximum value of
\begin{equation}
\left|2k_F{\partial  P_S^{\rm 2D}({\bf  q},\omega) \over
\partial q}\bigg|_{2k_F}/P_S^{2D}(2{\bf
k_F},\omega)\right|\sim
\left({\epsilon_F\over
\Delta}\right)^{{1\over 2}}.
\label{remff2d}
\end{equation}
In one dimension the response-function magnitude reaches
a maximum of
\begin{equation}
\left|P_S^{\rm 1D}(2{\bf k_F},\omega)\right|\sim{N^*\over
4}\ln\left({\epsilon_F^2\over \Delta^2-(\hbar\omega/2)^2
}\right),\label{remff1d}
\end{equation}
whereas in the normal metal it diverges logarithmically:
\begin{equation}
 P_M^{\rm 1D}({\bf  q},\omega)= N^*\bigg\{
{1\over 4x}{\rm ln}\left|{(1-x)^2-\nu^2
\over (1+x)^2-\nu^2}\right| + {i\pi \over 4x
}[\theta(1-|x+\nu|)-\theta(1-|x-\nu|)]\bigg\}\label{1dpi}
\end{equation}

These results are quite similar in implication to Hurault's. However,
the phonon frequency regime $\hbar\omega < 2\Delta$ is unphysical in ordinary
superconductors and rare in high-temperature superconductors. We now
turn our attention to the more physical frequency regimes.

\subsection[]{$2\Delta\alt \hbar\omega$}

When the phonon energy exceeds the excitation gap, the
superconductor recovers an anomaly.
For $q>2k_F$ the minimum energy quasiparticle mode creates
two quasiparticles with
momentum $q/2>k_F$.
Because both of these quasiparticles are created with a
momentum greater than the Fermi momentum,
this mode does not have an analogy in the
normal state.  The superconductor, therefore, has
lower energy excitations at high $q$ than
the normal metal, as can be seen in Fig. 1b.

For a fixed $\hbar\omega>2\Delta$ there are two regimes of $q$,
separated by the solid line in Figs. 1a and 1b.
For small $q$, ${\rm
Im}P_S({\bf q},\omega)\ne 0$ because the minimum excitation energy is less
than the driving frequency.  For large $q$ no
excitable modes of the electron gas exist, so ${\rm Im}P_S({\bf q},\omega)=0$.
Therefore, the imaginary part of $P_S({\bf q},\omega)$, and
by implication from the
Kramers-Kronig relations the real part as well, cannot
be analytic functions of $q$.  The momentum $q_c(\omega)$
beyond which no modes of frequency $\omega$ or less exist
is the anomaly's momentum.  A {\it nonspherical} Fermi
surface does not affect the analytic
form of the anomalies. If the Fermi surface
is known, the anomaly momenta in various directions can be calculated.
I will now derive the form of the nonanalyticity of $P_S({\bf q},\omega)$
at $q=q_c(\omega)$.

The nonanalyticity in $P_S({\bf q},\omega)$ at $q=q_c(\omega)$ can be
extracted by expanding the energy denominator in Eq. (\ref{Psc}) around the
anomaly's momentum:
\begin{equation}
E_{\bf k} + E_{{\bf q}-{\bf k}} = 2E_{{\bf
q_c}/ 2}+ {\epsilon_{{\bf q_c}/ 2}(p^2-q_c^2/4)+\epsilon_{{\bf
q_c}/ 2}(p^2+q^2+2pq{\rm cos}\theta-q_c^2/4)\over
2mE_{{\bf q_c}/2}/\hbar^2}. \label{expanded}
\end{equation}
This expansion is valid when the quantities in parenthesis
are small compared to $\hbar\omega/2$, which will usually mean
small compared to $\Delta$.  This expansion, therefore, is
only valid for states ${\bf k}$ and ${\bf q} - {\bf k}$ in
a region of dimension $(\hbar/\xi)$ around the momentum ${\bf q_c}/2$.
Consider the sum in Eq. (\ref{Psc}) to be restricted to
this region.  An evaluation of that sum, which will follow
and will be called $\tilde P_S({\bf q},\omega)$,
accurately gives ${\rm Im} P_S({\bf q},\omega)$ and the
nonanalytic part of ${\rm Re} P_S({\bf q},\omega) $ near
the nonanalyticity at $q=q_c$.

Since $2E_{{\bf q_c}/ 2}=\hbar\omega$ and the coherence
factor is smooth over a momentum $(\hbar/\xi)$, the sum can be
written as the following integrals in one, two and three
dimensions:
\begin{eqnarray}\tilde P_S^{\rm 1D}({\bf q},\omega) = -N^*{\hbar\omega\over
\epsilon_{\bf q_c}/2}&&{k_F\over 4}
C({{\bf q}\over 2},{{\bf q}\over 2})
\int_{{q_c\over 2}-a}^{{q_c\over 2}+a}
{dp \over 2p^2-q_c^2/2+q^2 - 2pq}
,\label{formw>2d1}\\
\tilde P_S^{\rm 2D}({\bf  q},\omega) =
-N^*{\hbar\omega\over
\epsilon_{{\bf q_c}/2}}&&{q_c\over 8\pi}
C({{\bf q}\over 2},{{\bf q}\over 2})\nonumber\\
&&\int_{{q_c\over 2}-a}^{{q_c\over 2}+a}
\int_{\pi-{a\over k_F}}^{\pi+{a\over k_F}}
{dpd\theta \over 2p^2-q_c^2/2+q^2 + 2pq{\rm cos}\theta},
\label{formw>2d2}\\
\tilde P_S^{\rm 3D}({\bf  q},\omega) = -N^*{\hbar\omega\over
\epsilon_{{\bf q_c}/2}}&&{q_c^2\over 8k_F}
C({{\bf q}\over 2},{{\bf q}\over 2})\nonumber\\&&
\int_{{q_c\over 2}-a}^{{q_c\over 2}+a}
\int_{\pi-{a\over k_F}}^{\pi}
{dpd{\rm cos}\theta \over 2p^2-q_c^2/2+q^2 + 2pq{\rm cos}\theta},
\label{formw>2d3}
\end{eqnarray}
where $a$ is a cutoff of order $\xi^{-1}$.

Evaluating the integrals above in the limit $q\to
q_c$ and defining $\tilde q=q-q_c$ yields the following
forms for the response functions:
\begin{eqnarray}
\tilde P_S^{\rm 1D}({\bf q},\omega) &&=
-iN^*{\hbar\omega\over \epsilon_{{\bf
q_c}/2}}
\left[ {\pi k_F \over 8\sqrt{-q_c\tilde q}}\right]
C({{\bf q}\over 2},{{\bf q}\over 2}),
\qquad\qquad\qquad \tilde q<0,\nonumber\\
&&= -N^*{\hbar\omega\over \epsilon_{{\bf
q_c}/2}}\left[{\pi
k_F\over 8\sqrt{q_c\tilde q}} \right]
C({{\bf q}\over 2},{{\bf q}\over 2}),
\qquad\qquad\qquad\ \ \  \tilde q>0, \label{1dg0}\\
\tilde P_S^{\rm 2D}({\bf  q},\omega) &&=
N^*{\hbar\omega\over
\epsilon_{{\bf q_c}/2}}{1\over 4}\left[
{\rm ln}\left|{\tilde q q_c\over 8a^2}\right| - i\pi\theta(-\tilde
q) \right] C({{\bf q}\over 2},{{\bf q}\over 2}),
\label{2dw>2d}\\
\tilde P_S^{\rm 3D}({\bf  q},\omega) &&= -N^*{\hbar\omega\over
\epsilon_{{\bf q_c}/2}}\left[P
+ i\pi\left({-\tilde qq_c\over 32k_F^2}\right)^{1\over 2}\right]
C({{\bf q}\over 2},{{\bf q}\over 2}),\qquad
\tilde q<0, \nonumber\\
&&= -N^*{\hbar\omega\over
\epsilon_{{\bf q_c}/2}}\left[P+
\pi\left({\tilde qq_c\over 32k_F^2}\right)^{1\over 2}\right]
C({{\bf q}\over 2},{{\bf q}\over 2}),
\ \qquad
\tilde q>0. \label{3dw>2d}
\end{eqnarray}
Here $\theta$ is the Heavyside step function and $P$ is an
uninteresting constant.
Only $\tilde P_S^{\rm 1D}(2{\bf k_F},\omega)$ has been reported
elsewhere\cite{Marsiglio}.
The change in form of the integrals in Eqs. (\ref{formw>2d1})-
(\ref{formw>2d3})
when $q$ passes through $q_c$ causes the nonanalyticities in
Eqs. (\ref{1dg0})-(\ref{3dw>2d}).
The forms of these nonanalyticities differ from those in the normal metal.

This primarily results from the different dispersion of the excitations
with momentum near $q_c/2$ between the metal and superconductor.
In a normal metal, for finite frequency, the anomaly's
momentum connects electronic states with different
velocities.  One electronic state rests on the Fermi surface and one does
not.
In the superconductor the two quasiparticle states have the same velocity,
causing an amplification of the density of states for
$E_{\bf k}+E_{{\bf q}-{\bf k}}=\hbar\omega$ and a
{\it stronger} nonanalyticity.
The prefactor in Eqs. (\ref{formw>2d1})-(\ref{formw>2d3}),
\begin{equation}
{\hbar\omega\over \epsilon_{\bf q_c/2}} =
{1\over 2}\left(1-\left[{2\Delta\over
\hbar\omega}\right]^2\right)^{-{1\over 2}},
\end{equation}
is due to the
square-root divergence near the Fermi surface
in the superconducting density of states.

\subsection{$\hbar\omega\gg 2\Delta$}

For large phonon frequencies, the anomaly's momentum exceeds
twice the Fermi momentum by well over $(\hbar/\xi)$.  In
this case, the small value of the coherence factor $C({{\bf
q_c}\over 2},{{\bf q_c}\over 2})$ renders the anomaly undetectable.  A
remnant of the normal metal's  Kohn anomaly still
exists.  Since all relevant excitations for this remnant
are near the
dashed line of Figs. 1ab, this situation is analogous to
finite temperature in a normal metal.  In the superconductor, phonons with
momenta on both sides of the dashed line
have zero-energy excitations, but their number
decreases markedly, over a momentum range $(\hbar/\xi)$,
upon crossing that dashed line.

\section{Experimental Implications}

The actual size of the phonon anomalies can be estimated
by including the response function in the phonon self energy in the
standard way\cite{Mahan} and then expanding about $\omega(2k_F) =
\omega_o$:
\begin{equation}
{\delta\omega({\bf q})\over \omega_o} = {\lambda\omega_o N^*\over 2}{\rm
Re}\left[{\tilde P_S({\bf q},\omega_o) -
P_M(2{\bf k_F},\omega_o)\over N^*}\right]
\label{rel}
\end{equation}
where $\lambda$ is the dimensionless electron-phonon coupling
constant. The linewidth is simpler to express:
\begin{equation}
{\gamma_S({\bf q},\omega_o)\over \gamma_o} = {{\rm Im}\tilde P_S({\bf
q},\omega_o)\over {\rm Im} P_M(2{\bf k_F},\omega_o)}.\label{ima}
\end{equation}
In comparing anomaly magnitudes I will assume that $\lambda$
in the superconductor equals the normal-metal  value. Recently\cite{Pickett}
the
average value of $\lambda$ over the whole Brillouin zone has been
calculated to be $1.37$.  The value of $N^*$ has been calculated\cite{Papa}
to be
$0.3/eV$. Fig.~\ref{lasr}
shows the real and imaginary part of the response function for a simple
model of $\rm La_{1.85}Sr_{.15}CuO_4$. The anomalies
evident in Fig.~\ref{lasr} are due to
crossing the pair threshold surfaces (shown for $\rm La_{1.85}Sr_{.15}CuO_4$ in
Fig.~\ref{thresh}). Fig.~\ref{phlasr} compares a phonon dispersion curve
in the normal and
superconducting state of $\rm La_{1.85}Sr_{.15}CuO_4$.
Clearly the anomaly should be larger
in superconducting $\rm La_{1.85}Sr_{.15}CuO_4$
than in metallic $\rm La_{1.85}Sr_{.15}CuO_4$. To show how the
two-dimensional anomalies are much larger than the three-dimensional
anomalies, Fig.~\ref{3dsc} shows the metallic and superconducting
response function for a three-dimensional spherical Fermi sea.

\section{Gap Anisotropy}

An anisotropic gap influences this anomaly only through the
coherence factor,
\begin{equation}
C({{\bf q_c}\over 2},{{\bf q_c}\over 2}) = \left(
{|\Delta_{{\bf
q_c}/ 2}|
\over E_{{\bf q_c}/ 2}}\right)^2,
\label{critcoh}
\end{equation}
which is independent of the phase of the gap.

So long as the gap is finite at $q_c/2$, the situation for anisotropic
gaps is essentially the same as for isotropic gaps.
A difference
occurs when the momentum spanning the Fermi surface connects two nodes
in the gap. For this situation the anomaly is weaker.
This has been analyzed numerically by
Marsiglio\cite{Marsiglio}.

\section{Conclusion}

The vanishing of the Kohn anomaly
for $\hbar\omega<2\Delta$ results from
the absence of electronic excitations at low energy in the
superconductor rather than from a smoothing of the Fermi surface.
The rapid decrease of the coherence factor of the minimum-energy
excitation for $\hbar\omega\gg 2\Delta$ also eliminates the Kohn
anomaly. A new regime exists when  $2\Delta\alt\hbar\omega$; here
superconductivity {\it enhances} the Kohn anomaly.
An appropriate material to examine when looking for this
effect would have phonon branches both above and below the
excitation gap at $q\sim 2k_F$, as well as a
quasi-two-dimensional electronic structure.  High-${\rm
T_c}$ superconductors like ${\rm La_{1.85}Sr_{.15}CuO_4}$
are such materials.

Since an extremely sensitive probe of surface
phonons exists in thermal-energy-inelastic-helium
scattering\cite{platinum}, I remark that similar arguments to
those presented in this paper may apply to surface phonons.

The anomalies discussed in this Article complete the catalogue of
pair-production threshold anomalies, begun by Bobetic\cite{Bobetic}
 and Schuster,\cite{Schuster} and elaborated for two-dimensional
superconductors recently by this author\cite{Flatte}.

\section*{Acknowledgments}

I would like to acknowledge useful conversations with W. Kohn, D.P.
Clougherty, and D.J. Scalapino. This work was supported by the National
Science Foundation under Grant No. PHY89-04035.

\begin{figure}
\caption{Minimum single-pair excitation energy
in a two or three-dimensional
superconductor (solid line) and normal
metal (dashed line). Here $\Delta/\epsilon_F = 0.01$.
 (a) Full range of momentum $q$. (b) Closeup of
the region near $q=2k_F$.\label{min}}
\end{figure}

\begin{figure}
\caption{Four possible extremal excitations for a fixed $\omega$. The
two dashed lines are energy contour surfaces at $\hbar\omega$ below and
above the Fermi surface, indicated by the solid circle. For momenta
smaller than (1) and (2) there are no real single-pair
excitations of the Fermi sea of this type. For momenta greater than (3)
and (4) there are no single-pair excitations of this type.\label{types}}
\end{figure}

\begin{figure}
\caption{$- P_M^{\rm 3D}({\bf  q},\omega)/N^*$ for
$\hbar\omega/\epsilon_F=0$ (solid line), $0.1$
(dashed line), and $0.2$ (dotted line). (a) Real part. (b) Imaginary part.
The solid line is not visible in (b) because ${\rm Im}P_M(q,0) = 0$.
\label{3dmetal}}
\end{figure}

\begin{figure}
\caption[]{$-P_M^{\rm 2D}({\bf  q},\omega)/N^*$ for
$\hbar\omega/\epsilon_F=0$ (solid line), $0.1$ (dashed line),
and $0.2$ (dotted line). (a) Real part. (b) Imaginary
part.
\label{2dmetal}}
\end{figure}

\begin{figure}
\caption[]{Kohn anomalies for phonons with momenta parallel to the (100)
direction in $\rm La_{1.85}Sr_{.15}CuO_4$.
Points on the dashed line correspond to Kohn anomalies when
phonon curves cross them. Solid lines are the low-energy phonons from
Ref. 16. The dotted line indicated the momentum of the
static anomaly.\label{100KA}}
\end{figure}

\begin{figure}
\caption[]{Same as Fig.~\ref{100KA} except for phonons with momenta
parallel to the (110) direction.\label{110KA}}
\end{figure}

\begin{figure}
\caption[]{Occupation number $f(\epsilon_{\bf k})$ for
a normal metal at $0K$ (dashed line) and the function $v_{\bf
k}$ for a superconductor at $0K$ (solid line).\label{vpfe}}
\end{figure}

\begin{figure}
\caption[]{The response function for a model of
$\rm La_{1.85}Sr_{.15}CuO_4$ with the correct curvature
at the points where a vector in the
(100) direction spans the Fermi surface. The Fermi
velocity was taken from Ref. 15. The frequency is fixed at
$18$~meV. The dashed line is for the normal metal and the solid line is
for the superconductor.  The nonanalyticities in these
curves
correspond to momenta where a phonon at this frequency would cross a
pair-production threshold surface shown in Fig.~\ref{thresh}. (a)
Real part. (b) Imaginary part.\label{lasr}}
\end{figure}

\begin{figure}
\caption[]{The pair-production threshold in the superconductor (dotted
line) in the
(100) direction is shown on
the same graph as the threshold in the normal metal (dashed line,
previously shown in
Fig.~\ref{100KA}). The gap magnitude is taken to be $7.5$ meV. The
four-pointed star indicates the momentum and energy of the superconductor's
anomaly. The five-pointed star indicates the momentum and energy of the
normal metal's anomaly.\label{thresh}}
\end{figure}

\begin{figure}
\caption{A phonon dispersion curve near $18$~meV in the normal and
superconducting state of $\rm La_{1.85}Sr_{.15}CuO_4$. (a) Frequency. (b)
Relative lifetime.\label{phlasr}}
\end{figure}

\begin{figure}
\caption{Response function for a superconductor (solid line) and normal
metal (dashed line)
with a three-dimensional Fermi surface. $\hbar\omega/\epsilon_F = 0.1$
and $\hbar\omega/2\Delta = 1.25$. (a) Real part. (b) Imaginary
part.\label{3dsc}}
\end{figure}

\end{document}